\documentclass[aps, prl, 
twocolumn,showpacs]{revtex4}
\usepackage[english]{babel}
\usepackage{graphics}
\usepackage{graphicx}
\usepackage{dcolumn}% Align table columns on decimal point
\usepackage{bm}% bold math
\usepackage{color}
\begin{document}

\title{Doping dependence of the mass enhancement in (Pb,Bi)$_2$Sr$_2$CaCu$_2$O$_8$ at the antinodal point
in the superconducting and normal state}
\author{T.K.\ Kim,$^1$ A.A.\ Kordyuk,$^{1,3}$ S.V.\ Borisenko,$^1$ A.\ Koitzsch,$^1$
M.\ Knupfer,$^1$ H.\ Berger,$^2$  J.\ Fink$^1$} 
\affiliation{$^1$ Leibniz-Institute for Solid
State and Materials Research Dresden, P.O.Box 270116, D-01171
Dresden, Germany\\
$^2$Institut de Physique de la Mati\`{e}re Complex, Ecole Politechnique F\'{e}d\'{e}rale de Lausanne, CH-1015 Lausanne, Switzerland\\
$^3$Institute for Metal Physics of the National Academy of Sciences of Ukraine, 03142 Kyiv, Ukraine}

\date{\today}

\begin{abstract}
Angle-resolved photoemission spectroscopy (ARPES) is used to study the mass renormalization of the
charge carriers in the high-T$_c$ superconductor (Pb,Bi)$_2$Sr$_2$CaCu$_2$O$_8$ in the vicinity of the
$(\pi,0)$ point in the superconducting and the normal state. Using matrix element effects at different photon
energies and due to a high momentum and energy resolution the bonding and the antibonding bands could
be separated in the whole dopant range. A huge anisotropic coupling to a bosonic collective mode is observed
below T$_c$ for both bands in particular for the underdoped case. Above T$_c$, the more isotropic coupling
to a continuum or a mode at much higher energy is significantly weaker.

\end{abstract}
\pacs{ 74.25.Jb, 74.72.-h, 79.60.-i }

\maketitle

Sixteen years after the discovery of high-T$_c$ superconductors (HTSCs) by Bednorz and Müller \cite{Muller} the 
understanding of the electronic transport in these systems is still full of unresolved issues. 
This holds both for the normal and for the superconducting state. In particular, it is not clear whether in 
these compounds which are close to an antiferromagnetic insulator the interaction between the electrons  is large 
enough to destroy coherence, i.e., the existence of well defined quasi-particles, and to result in
a new state of matter \cite{Anderson}. On the other hand, one can possibly 
analyze the  available experimental data in terms of a strong 
coupling to bosons such as phonons or spinfluctuations, leading to coherent and incoherent states.
Moreover, there are intermediate scenarios \cite{Varma}.
Therefore it is interesting to investigate in which
dopant and temperature range coherent states are present.

Traditional angle-resolved photoemission spectroscopy (ARPES) has been an important experimental probe to map
out the bandstructure of solids. Recent improvements of the energy and the momentum resolution opened
the possibility to study the many-body properties of solids. The deviation from the 
independent particle dispersion close to the Fermi level gives information on the renormalization of bands or
on the enhanced effective mass. The width of the peaks reflects the scattering rate or the
inverse lifetime of the photohole which should behave in a similar way as the charge carriers in a p-type doped 
HTSC. Both, the mass renormalization and the finite width of the spectral weight are normally described in terms
of the real and the imaginary part of the self energy $\Sigma$, respectively.

Numerous ARPES studies on HTSCs \cite{Damascelli,Campuz} have shown that the renormalization 
is strongly momentum dependent, i.e., it is a function of 
the position on the Fermi surface. Up to now there are many ARPES studies along the nodal direction 
(the $\Gamma-(\pi,\pi)$ line) where in the 
superconducting state no gap is opened. All these studies indicate that there is a mass 
renormalization m*/m=1+$\lambda$ at the Fermi level by a factor of about 2 corresponding 
to a coupling constant $\lambda$ of about 1.
The renormalization extends to an energy of about 70 meV below the Fermi level indicated by a kink in 
the dispersion. Currently there is no consensus on the origin of this renormalization. 
Coupling to collective bosonic modes such as phonons or spin fluctuations are discussed.

At the antinodal point (the $(\pi,0)-(\pi,\pi)$ Fermi surface crossing)
where the superconducting gap has a maximum, only a few experimental studies exist \cite{Damascelli,Campuz,
Kaminski,Dessau}. 
Moreover, the situation is more complicated by the fact that in the drosophila of the ARPES studies on HTSCs,  
Bi$_2$Sr$_2$CaCu$_2$O$_8$ (Bi2212), the so-called bilayer splitting
into a bonding (B) and an antibonding (A) band appears in this region of the Brillouin zone 
due to an interaction between two adjacent CuO$_2$ layers.  This bilayer splitting could not be resolved in many
previous low resolution experiments and therefore the existence of coherent states could be overlooked. 
On the other hand,
the investigation of bilayer systems offers the unique possibility to study the interaction of CuO$_2$ layers which is
of extreme importance for the understanding of the mechanism of high-T$_c$ superconductivity,
since T$_c$ increases with the number of interacting layers up to 3 and then decreases again.

In this Letter we present high resolution ARPES data of Bi2212 around the $(\pi,0)$ point for the whole 
dopant range from the overdoped (OD) via optimally doped (OP) to underdoped (UD) samples. Contrary to 
previous experiments
\cite{Kaminski}, we find dispersing renormalized bands not only in the superconducting state 
but also in the normal state in the whole dopant range, even for UD samples. 
Since we used the photon energy dependence of the excitations from the B and A bands 
we could determine the respective renormalization of both bands. It turned out that above T$_c$ 
the renormalization is rather isotropic and 
weaker than in the superconducting state. No kink and no specific energy scale could be detected. In the 
superconducting state the renormalization due to a coupling to a collective bosonic mode is very strong and 
increases with decreasing dopant concentration. The results are in contrast to the traditional view, 
in which due to correlation effects in the
normal state of UD samples no dispersing states exist and in the superconducting state the number of 
coherent states are expected to increase. Thus, the present results indicate a continuous transition from a more
conventional metal to a strange metal with a small number of coherent states and not a transition into a new state 
of matter when going from the OD to the UD region.

The ARPES experiments were carried out at the BESSY synchrotron radiation facility using the U125/1-PGM
beam line and a SCIENTA SES100 analyser. Spectra were taken with various photon energies ranging
from 17 to 65 eV. The total energy resolution ranged from 8 meV (FWHM) at photon
energies h$\nu$=17-25 eV to 22.5 meV at h$\nu$=65 eV. The momentum resolution was set to 0.01 {\AA}$^{-1}$ 
parallel to the
$(\pi,0)-(\pi, \pi)$ direction and 0.02 {\AA}$^{-1}$ parallel to the $\Gamma-(\pi,0)$ direction. Here we focus
on spectra taken with
photon energies of 38 eV and 50 (or 55 eV) to discriminate between B and A bands.
The polarization of the radiation was along the $\Gamma-(\pi,0)$ direction. Measurements have been performed on
(1$\times$5) superstructure-free Pb-Bi2212 single crystals ranging from OD 
(OD69, T$_c$= 69 K) via OP (OP89K, T$_c$= 89 K) to 
UD(UD77, T$_c$= 77 K) samples. The dopant concentration was calculated from the empirical equation of T$_c$ vs. the 
dopant concentration \cite{Tallon}. 

\begin{figure}%[b!]
\includegraphics[width=7.5 cm]{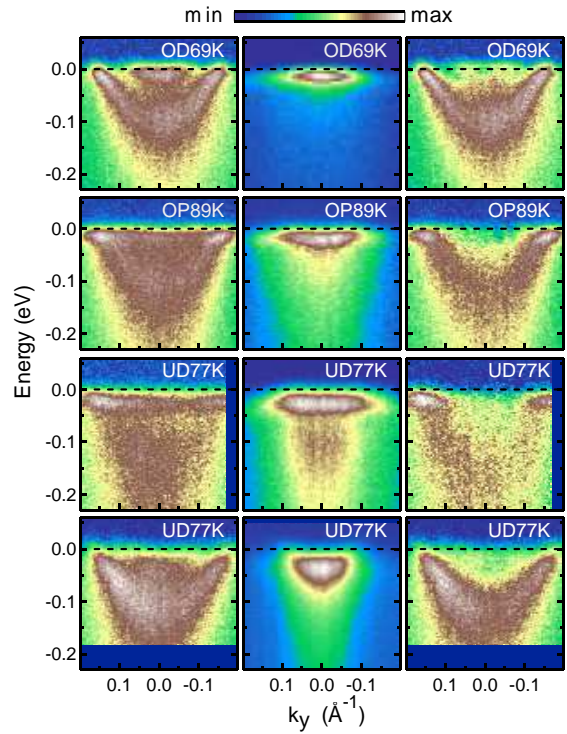}
\caption{ARPES intensity plots as a function of energy and wave vectors along the
$(\pi,\pi)-(\-pi,\pi)$ direction of overdoped (OD), optimally doped (OP) and underdoped  (UD) Pb-Bi2212 
superconductors taken at T = 30 K (upper 3 rows). Zero  corresponds to the $(\pi,0)$ point.
Fourth row: data for an UD sample taken at T = 120 K. 
Left column: data taken with a photon energy  h$\nu$=38 eV, at which the signal from the bonding band is maximal.
Middle column: data taken at h$\nu$=50 eV (or 55 eV), where the signal from the antibonding 
band is dominant. Right column: subtraction of the latter from the former yielding the spectral
weight of the bonding band (see text for details).} \label{EDMs}
\end{figure}

In Fig.~\ref{EDMs} we show a collection of data 
for wave vectors close to the 
$(\pi,\pi)-(\pi,-\pi)$ line, centered around the $(\pi,0)$ point. As shown previously
\cite{Kordyuk,Borisenko} and supported by theoretical calculations \cite{Bansil}, 
due to matrix element effects the data taken with h$\nu$= 38 eV represent mainly
the B band with some contributions from the antibonding band while the data taken with h$\nu$= 50 (or 55) eV 
have almost pure A character. The subtraction of the latter from the h$\nu$= 38 eV data represents almost
the pure B bands. This was performed by scaling the h$\nu$=50 eV spectral weight in such a way
that at the $(\pi,0)$ point no negative intensity occurs. Using this procedure one clearly recognizes 
that in the whole doping range and even
in the UD samples the B and the A band can be well separated.

In the superconducting state these data show  strong changes 
upon reducing the dopant concentration:
the B and most clearly seen the A band moves further and further below the Fermi level and
the renormalization strongly increases as indicated
by the appearance of a flat dispersion below the gap energy. At $\sim$70 meV  for the B band a kink and a reduction 
of spectral weight is observed, both increasing with decreasing
dopant concentration. 

\begin{figure}%[b!]
\includegraphics[width=7.5 cm]{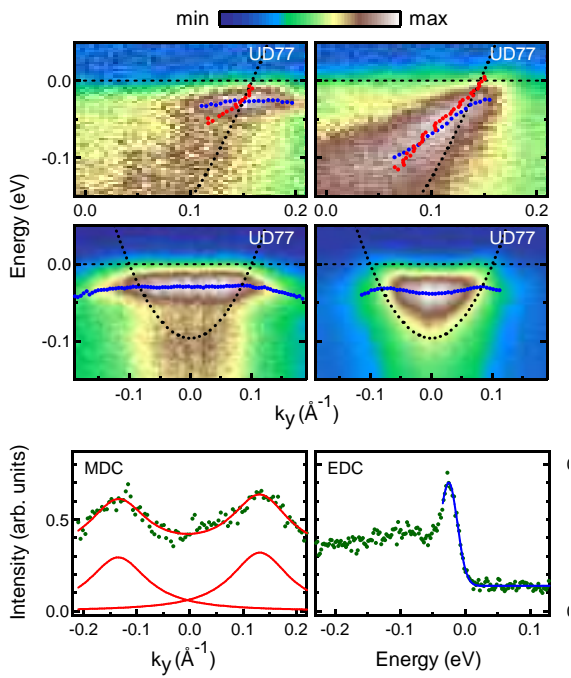}
\caption{First row: dispersion of the bonding band near the antinodal point in an underdoped  Pb-Bi2212 
sample at T=30 K (left) and at T = 120 K
(right) as derived from
EDCs (blue points) and MDCs (red points) together with the bare electron dispersion (black points).
Second row: the same data for the antibonding band.
Third row: MDCs (left panel) at 30 meV and EDCs (right panel) of the bonding band
at $k_F$ including fits to obtain the maximum.} 
\label{EDCMDC}
\end{figure}

In Fig.~\ref{EDCMDC} we show on an enlarged scale data for the B and A band of the UD77 sample. 
Also included are dispersions derived
from constant energy scans or momentum distribution curves (MDCs) in red for the B band and from
constant momentum scans or energy distribution curves (EDCs) in blue for both bands. 
As previously discussed \cite{Campuz,Dessau}
the EDC and the MDC dispersions are quite different for weakly dispersing states, i.e., in the superconducting state, while
they are similar in the normal state. Below T$_c$ the EDC derived dispersion shows the typical BCS-type back-dispersion
due to particle-hole mixing. At 120 K a pseudogap is observed and there is no observable back-dispersion.
This could explain the absence of a coherent peak in the tunnelling spectroscopy data in the pseudogap
regime \cite{Fischer}. 

\begin{figure}%[t!]
\includegraphics[width=7.5 cm]{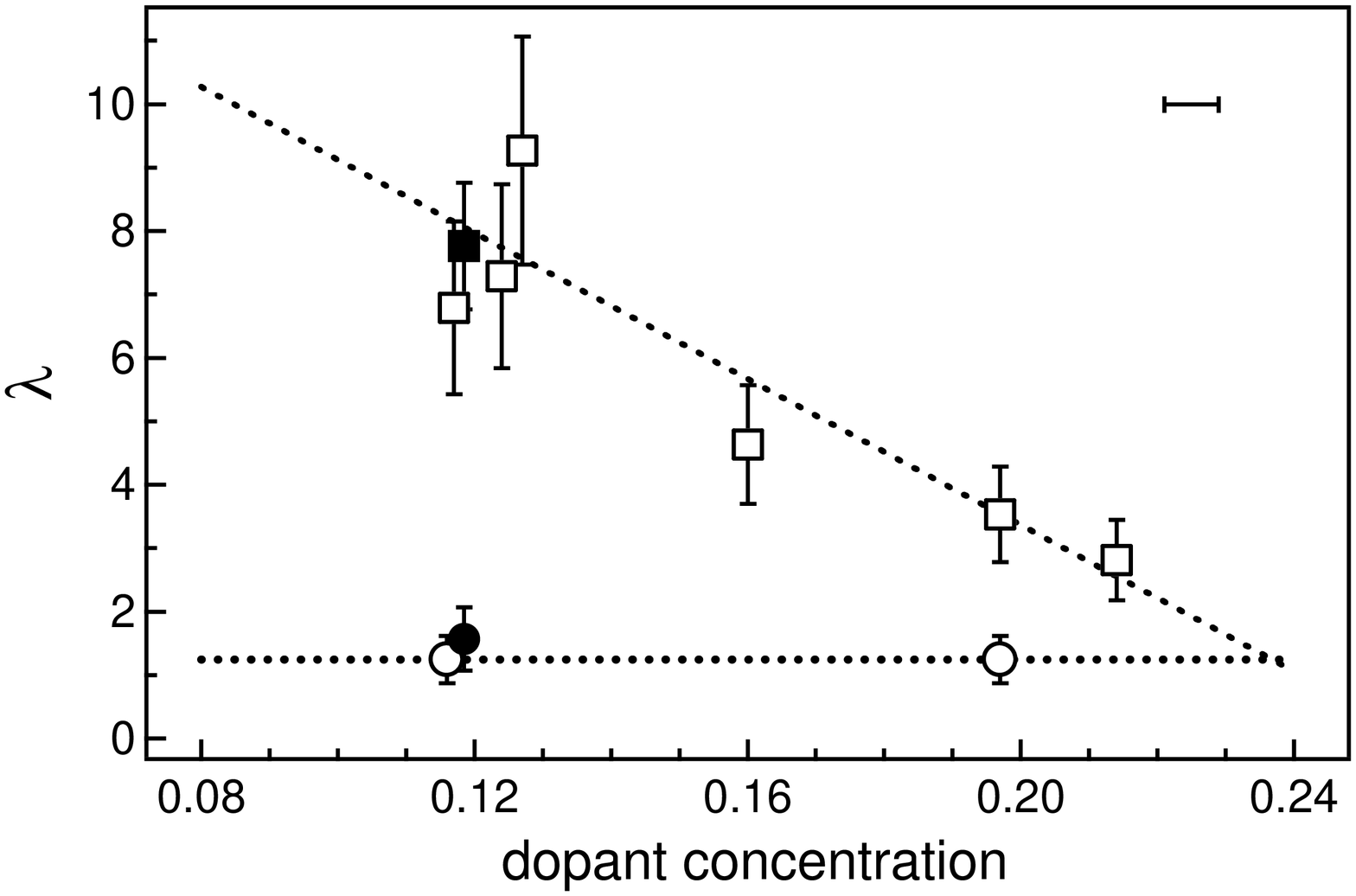}
\caption{The coupling strength parameter $\lambda$ as a function of dopant concentration. 
Squares: superconducting state, circles: normal state.
Open (solid) symbols: bonding (antibonding) band. The dashed lines are guides to the eye.
Horizontal bar: experimental error in the dopant concentration.} 
\label{lambda}
\end{figure}

The EDC depicted in Fig.~\ref{EDCMDC}, which was measured in the superconducting state at k$_F$ of the B band,
is typical of a strong coupling of the charge carriers to a bosonic mode \cite{Damascelli,Campuz,Dessau}.
For this sample, the dip-energy, ${E_D=\Delta+E_M}$, occurs at  70 meV which with a superconducting gap, 
$\Delta$= 30 meV, corresponds to
a mode energy $E_M$ of 40 meV \cite{Eschrig}. Then the 
dispersion E(k) in the vicinity of the gap can be approximated by the expression
E(k) =$(\epsilon_k ^2 + \Delta ^2)^{1/2}$
where $\epsilon_k$ is a renormalized band
dispersion which is approximated by a linear dispersion ${\epsilon_k=v_F^b(k-k_F)/(1+\lambda)}$ where $v_F^b$ 
is the Fermi velocity of the {\it bare} particles and $k_F$ is the Fermi wave vector \cite{Schrieffer}.
The bare Fermi velocities were taken from a previous 
evaluation of the Fermi surface and data along the nodal direction in terms of a tight-binding
bandstructure (also shown in Fig.~\ref{EDCMDC}) \cite{Kordyuk_gap_ARPES}. 
This bandstructure is close to that derived from an analysis of the anisotropic
plasmon dispersion \cite{Nucker} and to that obtained from LDA bandstructure calculations \cite{Andersen}.
The values for $\lambda$, presented in Fig.~\ref{lambda}, and $\Delta$ values (UD: 30 meV; OP: 22 meV; OD: 16 meV)
were obtained by fits of E(k)
to the EDC dispersions. In OP and OD samples, where steeper dispersions are observed, similar values of $\lambda$
were also derived from MDC dispersions. For the A band reliable values for $\lambda$ could be only obtained
for an UD sample since this band is just crossing the Fermi level. 

With decreasing dopant concentration there is a 
strong increase of the coupling strength from $\lambda$
of about 3 to huge values of about 8. We emphasize that the increase seems to be continuous 
and there is no break between the OD and the UD region. For the A band in
the UD77 sample a similar huge mass renormalization has been detected (see Fig.~\ref{lambda}).
The results for the A and the B band in the superconducting state can also be summarized 
in a different way. With decreasing dopant concentration the coherence factor
Z=1/(1+$\lambda$) decreases from 0.25 to 0.1. This means that for UD samples below T$_c$ only 
about 1/9 of the spectral  weight near E$_F$  represent coherent states.

Above T$_c$ we see dispersive states even in UD samples and
the renormalization of both bands is strongly reduced (see Fig.~\ref{EDCMDC}).
There is also no kink in the MDC dispersion at $\sim$70 meV
due to a coupling to a bosonic mode. The mode is either at much higher energies or there
is a continuum which leads to the renormalization. The coupling constant as derived
from the differences in the slopes of the 
measured dispersion and the bare particle dispersion is close to one,
independent of the dopant concentration. These values are also depicted in Fig.~\ref{lambda}. Comparing these values
with similar $\lambda$ values at the nodal point indicates that the renormalization in the normal
state is rather isotropic along the Fermi surface.

\begin{figure}%[t!]
\includegraphics[width=7.5 cm]{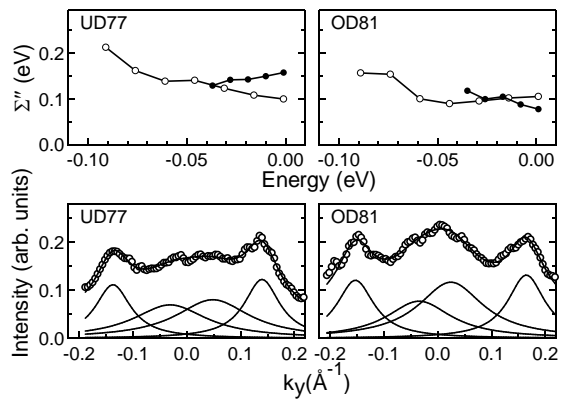}
\caption{Upper row: imaginary part of self energy
$\Sigma$'' as derived from MDC fits. 
Open (solid) symbols: bonding (antibonding) band.
Second row: MDC fits for OD and UD samples at energies of 15 and 35 meV, respectively.} 
\label{width}
\end{figure}

To obtain further information on the renormalization of the dynamics of the charge carriers in the normal state, 
we have evaluated
the MDC width of the A and the B bands which multiplied by $v_F^b$, is a measure of the imaginary part
of $\Sigma$  plus contributions due to the finite energy and momentum resolution. 
Typical MDC cuts measured with a photon energy of 38 eV are shown  in Fig.~\ref{width}
for UD and OD samples.  These cuts can be well fitted by 4 Lorentzians,
2 corresponding to the B and 2 others corresponding to the A band. For the OD  and the UD samples the
derived scattering rates at E$_F$ are almost the same for both  the B and the  A bands. 
Values between 80 and 160 meV are derived for the antinodal point at T= 120 K. 
These values are not far from our 
values ($\sim$ 100 meV) derived at the same
temperature and at the same energy at the nodal point. Moreover, at energies smaller than about 100 meV 
the scattering rates are only slightly
dependent on energy (see Fig.~\ref{width}). This is an indication that the modes to which the charge carriers are coupled, 
are at higher energies or
form a continuous spectrum.

The results obtained can  be summarized in the following way: in the normal state a rather 
isotropic mass renormalization of the electronic states without a clear energy scale
is observed. The strength of the coupling corresponds to $\lambda$
$\sim$1. Below T$_c$ and in particular in the UD samples a much higher strongly anisotropic 
coupling to a collective bosonic mode could be detected.
We emphasize that the difference in the mass renormalization at the nodal point between the superconducting 
and the normal state is rather small (less than 5 \%) .

It is difficult to interpret the entirety of our results  in terms of a conventional isotropic coupling to phonons. A probably 
more promising model would be
a coupling to spin fluctuations \cite{Chubukov}. Although this scenario has been applied in many previous  
ARPES studies the present work
provides a much more detailed picture. Above T$_c$ the system starts from charge carriers 
which are coupled to a continuum of spin fluctuations.
The strength of the coupling is not strongly dopant dependent. Both the A and the B band feel a similar 
coupling as derived from the mass renormalization and from the scattering rates. 
Previous ARPES studies at the
$(\pi,0)$ point could not resolve the bilayer splitting and could not follow the flat dispersion of bands 
and therefore came to the conclusion that in the UD region above T$_c$ there are only incoherent states. 
In the present study, renormalized dispersive and possibly coherent states are even detected
above T$_c$ in UD samples. 

In the spin fluctuation scenario,
below T$_c$, the opening of the gap leads via a feed-back process to a magnetic resonance mode at E$_M$
detected by inelastic neutron scattering \cite{neutron} 
to which the charge carriers couple.
Due to the fact that the bilayer splitting of the band could be resolved, a quantitative analysis of the 
coupling strength could be performed for  the B band in the entire doping range and for the A band for an UD sample.
In a  recent theoretical work \cite{Eschrig} it was pointed out that according to magnetic
susceptibility measurements using inelastic neutron scattering the magnetic resonance mode couples the
A band only to the B band and vice versa. There is no coupling via the resonance mode within a band. In that
paper \cite{Eschrig} the authors came to the conclusion that in the OD region the mass enhancement of the B band
should be large since the van Hove singularity of the A band comes close to the Fermi level and leads to a large
susceptibility ${\chi_{AB}}$. This is at variance with our experimental results where $\lambda$ for the B band 
does not show a maximum in the OD region. On the other hand, it is remarkable that the coupling of the B band
to the resonance mode starts when the A band crosses the Fermi level (in the OD region). Moreover,
the result that in the UD region $\lambda$
is similar for the both bands is understandable, since the Fermi velocities and therefore the density of states
and the odd susceptibilities ${\chi_{AB}}$ and ${\chi_{BA}}$ should be comparable.

This work was financially supported in part by the Schweizerische Nationalfonds zur Foerderung der 
Wissenschaftlichen Forschung.

\end{document}